\documentclass[aps,prb,twocolumn,showpacs,superscriptaddress]{revtex4-1}
\usepackage{latexsym}
\usepackage{amssymb}
\usepackage{graphicx}
\usepackage{amsmath}
\usepackage{bm}
\usepackage[colorlinks,
          linkcolor=black,
            citecolor=black,
            urlcolor=blue
           ]{hyperref}
\usepackage{verbatim}
\usepackage{mathrsfs}
\usepackage{extarrows}
\usepackage{comment}

\newcommand{\scrH}{{\mathscr{H}}}

\begin{document}

\title{Theory of electron spin resonance in one-dimensional topological insulators with spin-orbit couplings: Detection of edge states}

\author{Yuan Yao}
\email{smartyao@issp.u-tokyo.ac.jp}
\affiliation{Institute for Solid State Physics, University of Tokyo, Kashiwa, Chiba 277-8581, Japan}
\author{Masahiro Sato}
\affiliation{Department of Physics, Ibaraki University, Mito, Ibaraki 310-8512, Japan}
\affiliation{Spin Quantum Rectification Project, ERATO, Japan Science and Technology Agency, Sendai 980-8577, Japan}
\author{Tetsuya Nakamura}
\affiliation{Department of Physics and Mathematics, Aoyama-Gakuin University, Sagamihara, Kanagawa 229-8558, Japan}
\author{Nobuo Furukawa}
\affiliation{Department of Physics and Mathematics, Aoyama-Gakuin University, Sagamihara, Kanagawa 229-8558, Japan}
\author{Masaki Oshikawa}
\affiliation{Institute for Solid State Physics, University of Tokyo, Kashiwa, Chiba 277-8581, Japan}

\begin{abstract}
Edge/surface states often appear in a topologically nontrivial
phase when the system has a boundary.
The edge state of a one-dimensional topological insulator is one of the
simplest examples.
Electron spin resonance (ESR) is an ideal probe to detect and analyze
the edge state for its high sensitivity and precision.
We consider ESR of the edge state of a generalized Su-Schrieffer-Heeger
model with a next-nearest neighbor (NNN) hopping and a staggered
spin-orbit coupling.
The spin-orbit coupling is generally expected to bring about
nontrivial changes on the ESR spectrum.
Nevertheless, in the absence of the NNN hoppings, we find that
the ESR spectrum is unaffected by the spin-orbit coupling
thanks to the chiral symmetry.
In the presence of both the NNN hopping and the spin-orbit coupling,
on the other hand, the edge ESR spectrum exhibits a nontrivial
frequency shift.
We derive an explicit analytical formula
for the ESR shift in the second-order perturbation theory, which agrees very well with a
non-perturbative numerical calculation.
\end{abstract}

\maketitle

\section{Introduction}

In recent decades, topological phases have become a central issue in
condensed matter physics.
An important class of topological phases is
topological insulators and topological
superconductors \cite{Hasan:2010aa,Qi:2011aa,Bernevig:2013aa,SQ_Shen-TIbook}.

In condensed matter and statistical physics,
one-dimensional (1-D) systems, which are amenable to several
powerful analytical and numerical methods,
often provide useful insights.
1-D topological phases are no exceptions.
One of the simplest 1-D models
possessing nontrivial topological nature is the Su-Schrieffer-Heeger
(SSH) model \cite{Su:1979aa}, which has been used to describe the lattice
structure of polyacetylene $[\text{C}_2\text{H}_2]_\text{n}$. The SSH model
can be also applied to the 1-D charge density wave systems, such as
quasi-one-dimensional conductors like TTF-TCNQ
(tetrathiofulvalinium-tetracyanoquinodime- thanide) and KCP
(potassium-tetracyanoplatinate) \cite{Schulz:1978aa}.
While the SSH model had been studied intensively much earlier than
the notion of topological phases {was} conceived, there is a renewed interest
from the viewpoint of topology.
In fact, distinct phases of the SSH model are classified
by the Zak phase~\cite{Zak:1989aa}
which is a topological invariant, and the bulk winding
number of the momentum-space Hamiltonian \cite{Asboth:2016aa}.
In this sense, the SSH model can be regarded as a 1-D topological
insulator.

An important nontrivial signature of many topological phases is
edge states.
The SSH model indeed possesses zero-energy edge states that
are protected by a chiral symmetry\cite{Asboth:2016aa}.
The number of edge states at a domain wall is
equal to the bulk winding number.
This is known as the bulk-boundary correspondence in the spinless
inversion-symmetric SSH model\cite{Asboth:2016aa}.
Experimentally, 1-D systems with boundaries or edges can be
realized by adding impurities to the material so that the
system is broken to many finite chains.
However, the edge states are often experimentally difficult to observe, since they are localized near the boundaries or the impurities
and their contribution to bulk physical quantities is small. 
Given this challenge, electron spin resonance (ESR) provides one of
the best methods to probe the edge states, thanks to
its high sensitivity.
In fact, the edge states of the $S=1$ Haldane chain were created by
doping impurities and then successfully
detected by ESR~\cite{Hagiwara:1990aa,Glarum:1991aa}.
Furthermore, combined with near-edge x-ray
absorption fine-structure experiments,
ESR was applied successfully to probe the magnetic edge
state in a graphene nanoribbon
sample~\cite{Joly:2010aa,Campos-Delgado:2008aa}.
Such a strategy could also be applied to 1-D topological insulators,
which are described by the SSH model.

Another intriguing nature of ESR is that it is highly sensitive
to magnetic anisotropies, such as the anisotropic exchange interaction,
single-spin anisotropy, and the Dzyaloshinskii-Moriya (DM) interaction.
The effect of magnetic anisotropies on ESR is well understood
only in limited circumstances, and there remain many open
issues~\cite{Oshikawa:1999aa,Oshikawa:2002aa}.
These magnetic anisotropies are often consequences of
spin-orbit (SO) coupling 
which generally breaks spin-rotation symmetry. 
{The effects of magnetic anisotropies and SO couplings also play important roles 
in magnetic dynamics in higher-dimensional topological phases~\cite{Hasan:2010aa,Qi:2011aa,Bernevig:2013aa,
SQ_Shen-TIbook,Shindou:2010aa,Nakamura:2016aa}. } 
Thus it is of great interest to study the effect of SO coupling
on ESR directly. However, this question has not been explored
in much detail so far.
An obstacle for the potential experimental ESR study of
SO coupling is the electromagnetic screening in
metallic systems.
This problem does not exist in insulators.
Unfortunately, band insulators are generally non-magnetic and
we cannot expect interesting ESR properties.
On the other hand, Mott insulators can have interesting magnetic
properties. However, strong correlation effects, which are
essential in Mott insulators, make theoretical analysis difficult.

In this context, the 1-D topological insulator provides a unique
opportunity to study the effects of SO coupling on ESR.
This would be of significant interest in several aspects.
Experimentally, the insulating nature makes the observation
of edge states by ESR easier.
Theoretically, the interesting effects of anisotropic
SO coupling on ESR can be studied accurately for the SSH model
of non-interacting electrons.
Moreover, the chiral symmetry, which is
essential for the well-defined topological insulator phase,
is often broken explicitly in realistic systems. When we introduce a chiral-symmetry breaking perturbation 
to the 1-D SSH model, the energy eigenvalues of the edge states generally 
deviate from zero energy. However, the edge states are expected to 
still survive and be localized near the edge 
if the perturbation is small enough. As we will demonstrate, the ESR of the edge state can detect 
the breaking of the chiral symmetry. 
The purpose of this paper is to present a theoretical analysis
on ESR of the edge states in 1-D topological insulators,
based on a generalized SSH model with SO couplings.
We demonstrate several interesting aspects of ESR, which will
hopefully stimulate corresponding experimental studies.

The paper is organized as follows. In Sec.~\ref{model}, we present the
model of interest and review the basic topological nature of 
the SSH model. The next three sections are the main part of 
this paper. The properties of edge states are discussed in detail 
in Sec.~\ref{edge}. 
In Sec.~\ref{sec:perturbation}, we obtain a compact analytical formula of 
the ESR frequency shift in perturbation theory with respect to 
SO coupling. Section~\ref{sec:numerical} is devoted to a direct numerical 
calculation of the ESR frequency shift, which is independent of 
the perturbative approach in Sec.~\ref{sec:perturbation}. We find that 
the perturbation theory agrees with the numerical results very well.
Finally, we present conclusions and future problems in Sec.~\ref{conclusion}.

\section{The model}
\label{model}

\subsection{A generalized SSH model}

First let us consider a generalized SSH model with SO coupling
\begin{eqnarray}
\label{H0}
\scrH_0&=&-\sum_{j=1}^{+\infty}\left\{
t \left[1+(-1)^j\delta_0\right]
c_{j+1}^\dagger
\exp{\left[i (-1)^j \frac{\phi}{2} \vec{n}\cdot\vec{\sigma} \right]} c_j
\right.\nonumber\\
&&\left.+\text{h.c.}\right\}
\end{eqnarray}
where
$c_j$ is the two-component electron annihilation operator
$c_j\equiv[c_{j\uparrow},c_{j\downarrow}]^T$ at the $j$-th site,
$t>0$ is the nearest-neighbor (NN) electron hopping amplitude,
and $-1\leq\delta_0\leq1$ is the bond-alternation parameter.
The angle $\phi$ and axis $\vec{n}$ (which is a unit vector)
parametrizes the SO coupling on the NN bond. 
The angle $\phi$ denotes the ratio of the SO coupling
to the hopping amplitude on the bond.
In this paper, we assume that $\phi$ is sufficiently small ($|\phi|\ll 1$), 
which is the case in many real materials.
Expanding $\phi$ to first order, we obtain a standard
form with so-called intrinsic and
Rashba SO couplings~\cite{Konschuh-TBSO_PRB2010}.

In our model Eq.~\eqref{H0},
SO coupling is assumed to be staggered along the chain.
This would be required, in the limit of $\delta_0=0$, if the
system had site-centered inversion symmetry.
In general, other forms of SO coupling including the
uniform one along the chain are also possible.
In this paper, however, we focus on the particular case
of the staggered SO coupling to demonstrate its interesting
effects on the ESR spectrum.

\subsection{The SSH model and its topological properties}

In the limit $\phi=0$, our model is reduced to the
standard SSH model
\begin{eqnarray}
\label{ssh}
\mathscr{H}_\text{SSH}=-\sum\left\{t
\left[ 1+(-1)^j\delta_0 \right]
c_{j+1}^\dagger c_j+\text{h.c.}\right\} .
\end{eqnarray}
Let us first consider a system of
$2N$ sites ($N$ unit cells) with the periodic boundary condition (PBC).
It is then natural to take the momentum representation
\begin{eqnarray}
\label{fouriera}
c_{2j,\sigma}&=&\frac{1}{\sqrt{N}}\sum_ka_{k,\sigma}
\exp{\left( ik(2j) \right)}, \\
\label{fourierb}
c_{2j+1,\sigma}&=&\frac{1}{\sqrt{N}}\sum_kb_{k,\sigma}
\exp{\left( ik(2j+1) \right)},
\end{eqnarray}
where the summation of $k$ is in the reduced Brillouin
zone $[0,\pi)$ with $k=n\pi/N$ and $n=0,\cdots,N-1$.
Corresponding to the each sublattice (even and odd), there are
two flavors of fermions, $a$ and $b$.
The Hamiltonian can then be written as
\begin{equation}
\mathscr{H}_\text{SSH}= \sum_k(a_k^\dagger,b_k^\dagger){h}_\text{SSH}(k)
\begin{pmatrix}
 a_k \\
 b_k
\end{pmatrix},
\label{eq.H_SSH}
\end{equation}
with 
\begin{equation}
{h}_\text{SSH}(k)\equiv d_x(k)\tau_x+d_y(k)\tau_y,
\end{equation}
where $\tau_{x,y,z}$ are Pauli matrices acting on the flavor space,
and $d_{x,y}(k)$ are real numbers
\begin{eqnarray}
d_x(k)=-2t\cos(k);\,\,d_y(k)=2t\sin(k)\delta_0 .
\end{eqnarray}
The spin indices are again suppressed in the Hamiltonian.

\begin{figure}[t]
\includegraphics[width=0.3\textwidth]{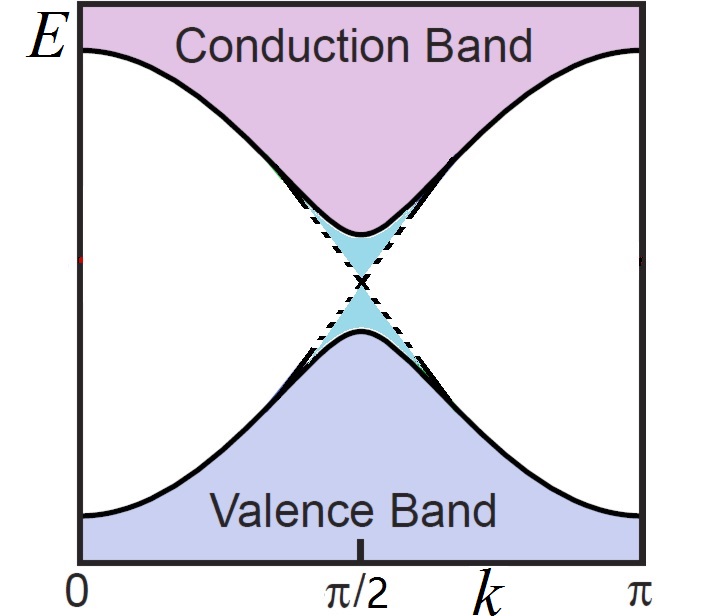}
  \caption{Band structure of the SSH model in Eq.~(\ref{ssh}) with a periodic 
boundary condition.~\cite{Hasan:2010aa} The solid and dashed lines 
respectively represent band structures of the insulating case 
with $\delta_0\neq0$ and the gapless point at $\delta_0=0$. 
The low-energy physics around $k=\pi/2$ can be described by 
one-dimensional Dirac fermion model.}
\label{indicationgap}
 \end{figure}

From this expression, the single-electron energy reads
\begin{equation}
 \epsilon(k) = \pm \sqrt{{d_x(k)}^2+{d_y(k)}^2}
= \pm 2t \sqrt{\cos^2{k}+ {\delta_0}^2\sin^2{k}} .
\end{equation}
The gap is closed at $k=\pi/2$ when $\delta_0=0$, while
the system has a gap $4t|\delta_0|$ whenever the bond alternation
does not vanish ($\delta_0 \neq 0$). 
The gapless point can be regarded as a quantum critical point
separating the two gapped phases, $\delta_0 < 0$ and $\delta_0 >0$.
Throughout this paper, we consider the half-filled case with
$2N$ electrons.
The bulk mode near this gap-closing point has a linear
dispersion relation indicated in Fig.~\ref{indicationgap}, and it can be
described by a one-dimensional Dirac fermion \cite{Bernevig:2013aa}.

It is evident from the Hamiltonian that each of the gapped
phases is simply a dimerized phase.
Nevertheless, we can identify them as a trivial insulator phase
and a ``topological insulator'' phase.
This can be understood by considering the system with
the open boundary condition.
Let us consider the chain of $2N$ sites labeled with $j=1,2,\ldots,2N$,
and the open ends at sites $j=1$ and $2N$.
For $\delta_0>0$ ($\delta_0<0$), sites $j=2n$ and $j=2n+1$
($j=2n-1$ and $j=2n$) form a dimerized pair, respectively.
As a consequence, for $\delta_0 >0$ the end sites $j=1$ and
$j=2N$ remain unpaired. The electrons at these unpaired sites
give rise to $S=1/2$ edge states.
In contrast, for $\delta_0 <0$, there are no unpaired sites
and thus no edge states.
In this sense, $\delta_0>0$ is a topological insulator phase
and $\delta_0<0$ is a trivial insulator phase.
Of course, considering the equivalence of the two phases
in the bulk,
such a distinction involves an arbitrariness.
That is, if we consider the an open chain of $N$ sites
with $j=0,1,\ldots,2N-1$, the edge states appear only for
$\delta_0<0$.
It is still useful to identify the gapless point at
$\delta_0=0$ as a quantum critical point separating
the topological insulator and the trivial insulator phase.

The particular shape of the Hamiltonian also implies the existence
of a chiral symmetry:
\begin{equation}
\{ {h}_\text{SSH}, \Gamma \} = 0,
\end{equation}
where $\Gamma \equiv \tau_z$.
The chiral symmetry turns out to be important for the
distinction of the two phases.
In the context of the general classification of topological insulators,
the present system corresponds to the
``AIII''  class with particle number conservation and the chiral symmetry
in one spatial dimension~\cite{Schnyder:2008aa, Chiu:2016aa}.

In a general one dimensional free fermion system,
we can define a topological invariant called the
Zak phase \cite{Zak:1989aa} for each band as follows:
\begin{eqnarray}
\gamma_\text{Zak}=i\oint_\text{BZ}\langle\Psi(k)|\triangledown_k|\Psi(k)\rangle,
\end{eqnarray}
where $|\Psi(k)\rangle$ is the Bloch wavefunction of the
band with the momentum $k$.
In the presence of the chiral symmetry,
$\gamma_\text{Zak}$ is quantized to integral multiples of $\pi$,
if the band is separated from others by gaps~\cite{Delplace-Zak_phase-2011}.

\begin{figure}
  \centering
\includegraphics[width=0.5\textwidth]{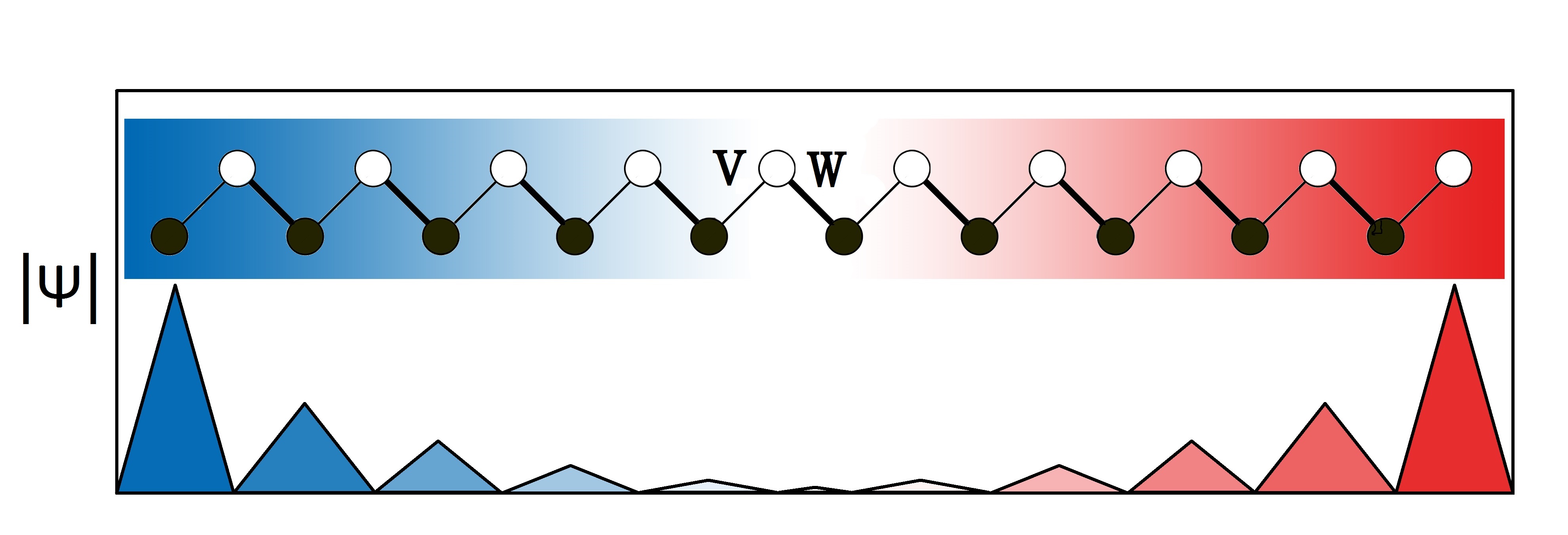}
\caption{Amplitude $|\psi|$ of the edge-state wave function of the SSH model in Eq.~(\ref{ssh}) under an open boundary condition.~\cite{Asboth:2016aa} 
Blue and red colors respectively represent the spatial distribution of 
the existing probability for left and right localized edge states. 
The total site number is set to be even, 
the dimerization parameter $\delta_0>0$, and symbols $v$ and $w$ denote 
hopping amplitudes $t(1-\delta_0)$ and $t(1+\delta_0)$, respectively. 
The wave-function amplitude decays exponentially into the 
bulk~\cite{SQ_Shen-TIbook}.}
\label{indicationamplitude}
 \end{figure}

For the present two-band SSH model in Eq.~\eqref{eq.H_SSH} with the chiral symmetry,
we can compute the Zak phase using the explicit Bloch wavefunction.
For the lower band,
\begin{equation}
|\Psi(k)\rangle= \frac{1}{\sqrt{2}}
\begin{pmatrix}
 \exp{(-i\phi_k)} \\
 1
\end{pmatrix},
\end{equation}
with $\phi_k\equiv\arctan[d_y(k)/d_x(k)]$. 
As a result, we find $\gamma_\text{Zak}/\pi=1$ for $0<\delta_0\leq 1$ 
in which the system is a topological insulator.
In the other case $-1\leq\delta_0<0$, where the system is a trivial
insulator, $\gamma_\text{Zak}/\pi=0$.
In this case, we can see that the Zak phase can be also
identifed~\cite{Delplace-Zak_phase-2011} with
a winding number of the Hamiltonian as
\begin{equation}
 \frac{\gamma_\text{Zak}}{\pi} = \frac{i}{\pi}
 \oint_\text{BZ}dk\triangledown_k\ln\left[d_x(k)-id_y(k)\right] .
\end{equation}
When $\gamma_\text{Zak}/\pi =1$,
there is an edge state localized at each end of an open finite chain
as shown in Fig.~\ref{indicationamplitude}.
This is the bulk-boundary correspondence~\cite{Ryu-Hatsugai2002}
in the SSH model.
In general, this topological number can take arbitrary integer
values, corresponding to $\mathbb{Z}$ classification of
BDI or AIII class in $d=1$ dimension.
However, in the present SSH model, its value is restricted to
$0$ or $1$.

\begin{figure}
  \centering
 \includegraphics[width=0.5\textwidth]{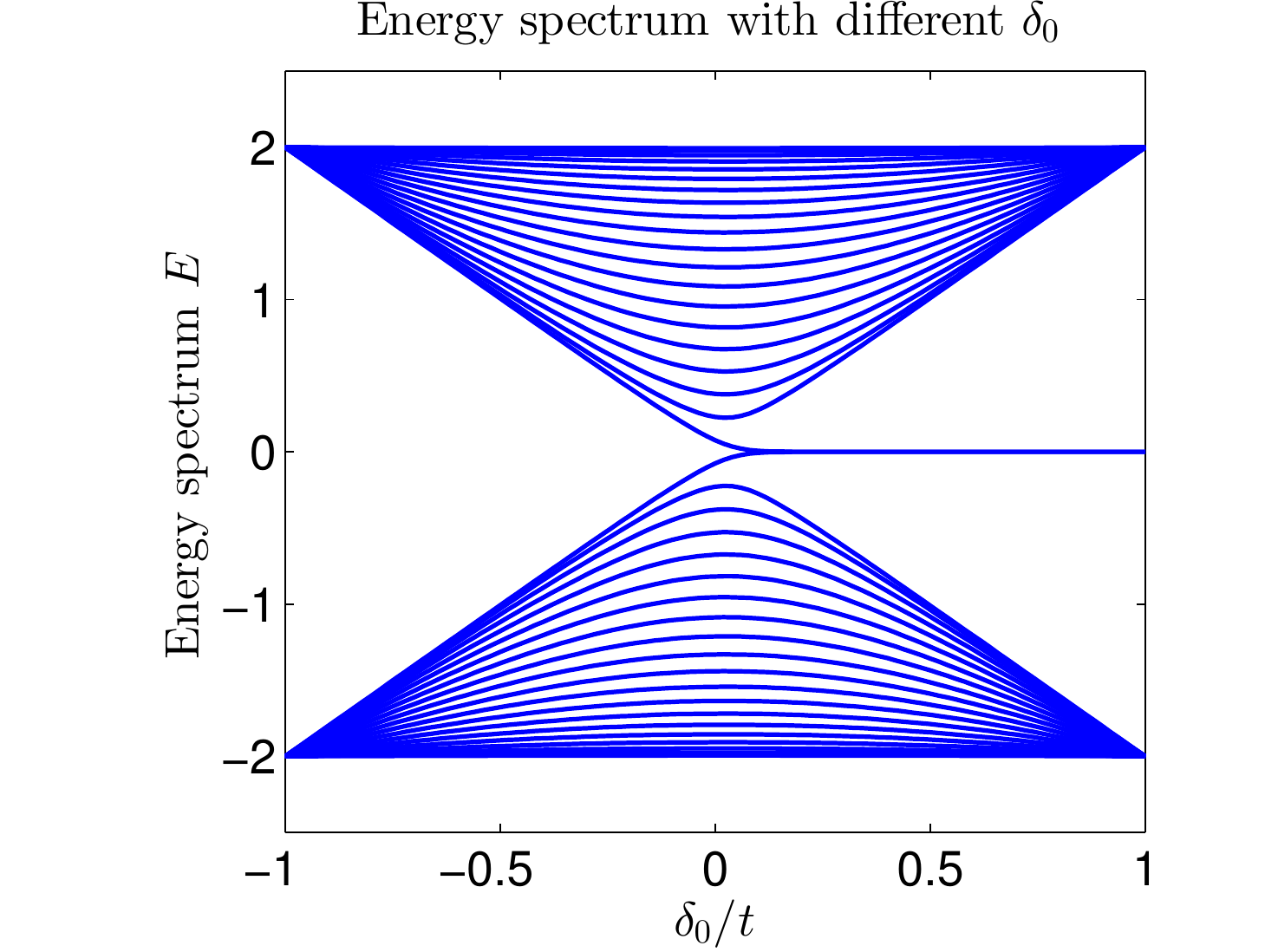}
  \caption{$\delta_0$ dependence of the energy spectrum of the SSH model 
with $t=1$ and $N=40$ under an open boundary condition. 
As $\delta_0\rightarrow0^+$, two localized edge states at the ends 
merge into the bulk. For the limit $N\rightarrow+\infty$, the edge states 
are strictly at zero energy as $0<\delta\leq t$ which are 
protected by the chiral symmetry.~\cite{Asboth:2016aa}}
\label{figedgebulk}
 \end{figure}

The existence of the edge states in the SSH model
can be demonstrated by an explicit calculation for a finite-size chain. 
In Fig.~\ref{figedgebulk}, we can see that, when $\delta_0$ decreases
from 1 to -1, the edge states merge into the bulk spectrum as
$\delta_0\rightarrow0^+$. In addition, when the thermodynamic limit,
$N\rightarrow+\infty$, is taken in the open-end SSH model, the edge
states are strictly at zero energy and topologically stable against any
local adiabatic deformation that
respects the chiral symmetry~\cite{Asboth:2016aa}.


\subsection{ESR of edge states}

Let us consider ESR of the 1-d half-filled topological insulator phases 
at the low-temperature and low-frequency limit $|\omega|,T\ll|\delta_0|t$ on which we focus in this paper. 
The ESR contribution from bulk excitations is negligible in this limit 
since there is a large bond-alternation driven band gap $4|\delta_0|t$. 
On the other hand, spin-1/2 edge states are located at the (nearly) zero energy point 
in the band-gap regime. Therefore, ESR is dominated by the edge state contribution. 

When the chiral symmetry is preserved and SO coupling
is absent, the edge spin is precisely equivalent to a free $S=1/2$.
In this case, the edge ESR spectrum is trivial, which means that it just consists of the delta function at the Zeeman energy.
However, breaking of the chiral symmetry and introduction of
SO couplings can bring a nontrivial change on the edge ESR spectrum.
In the following, we shall analyze this effect theoretically.

In ESR, absorption of an incoming electromagnetic wave
is measured under a static magnetic field.
Thus we introduce the Zeeman term for the static, uniform magnetic field
\begin{equation}
\mathscr{H}_Z =
-\frac{H}{2}\sum_{j=1}c^\dagger_j\left(\vec{\sigma}\cdot\vec{n}_H\right)c_j,
\end{equation}
where $\vec{n}_H$ is a unit vector representing the direction
of the  magnetic field,
$H>0$ is its magnitude, and
$\vec{\sigma} = (\sigma^x,\sigma^y,\sigma^z)$.

In the paramagnetic resonance of independent electron spins,
absorption occurs for the oscillating magnetic field
perpendicular to the static magnetic field, which is measured
in the standard Faraday configuration.
Therefore, in this paper, we assume that the oscillating magnetic
field is perpendicular to the static magnetic field $\vec{n}_H$. 
The frequency of the electromagnetic wave is denoted by $\omega$.

In an electron system with the SO interaction,
the electric current operator contains a ``SO current''
that involves the spin operator.
Since the electric current couples to the oscillating electric
field, in the actual setting of the ESR experiment,
the optical conductivity due to the SO current
also contributes to the absorption of the electromagnetic wave
with a spin flip.
This effect is called Electron Dipole Spin Resonance
(EDSR)~\cite{Rashba:1960aa, Rashba:2003aa, Efros:2006aa, Bolens:2017ab}.
The EDSR contribution is generically larger than ESR if SO coupling is at the same order as the Zeeman splitting \cite{Shekhter:2005aa,Maiti:2016aa}, as their relative contributions are of $(a/\lambda_\text{C})^2\approx10^6$, where $a\approx10^{-10}$m is the lattice spacing and $\lambda_\text{C}\approx10^{-13}$m the Compton length of the electron \cite{Bolens:2017ab}. 
In general, EDSR requires a separate consideration from
ESR as they involve different operators \cite{Bolens:2017ab}. 
Nevertheless, in the low-temperature/low-frequency regime,
only the two spin states of the edge state are involved.
Thus, although EDSR contributes to the absorption
intensity differently from ESR, the resonance frequency is identical
between the ESR and EDSR.
With this in mind, we do not consider EDSR explicitly
in the rest of the paper.
It should be noted that, for a higher temperature or a higher
frequency, the EDSR contribution to the absorption spectrum
is rather different from the ESR one, as the absorption spectrum involves bulk
excitations.

We now consider the ESR in the system with the Hamiltonian $\scrH_0+\scrH_z$.
Within the linear response theory~\cite{Kubo:1954aa},
the ESR spectrum is generally given by the
dynamical susceptibility function in the limit of
zero-momentum transfer
\begin{eqnarray}
\label{kubo}
\chi''_{+-}(q=0,\omega>0)&=&-\text{Im}\mathscr{G}^\text{R}_{+-}(q=0,\omega),
\end{eqnarray}
where
\begin{eqnarray}
\mathscr{G}^\text{R}_{+-}(0,\omega)&=&-i\int_0^{\infty}dt\sum_{r,r'}
\langle[s^+(r,t),s^-(r',0)]\rangle e^{i\omega t}\nonumber\\
&=&-i\int_0^{+\infty}dt\langle[S^+(t),S^-(0)]\rangle e^{i\omega t} ,
\end{eqnarray}
where
$S^\pm$ means the ladder operator defined
with respect to the direction $\vec{n}_H$ of the static field,
$\langle\cdots\rangle$ denotes the
quantum and ensemble average at the given temperature $T$,
and
\begin{equation}
S^{\pm}(t)\equiv\sum_{r}\exp(i\mathscr{H}t)s^{\pm}(r)\exp(-i\mathscr{H}t)=\sum_{r}s^{\pm}(r,t),
\end{equation}
{where $\mathscr{H}$ is the static Hamiltonian we consider 
(e.g., $\mathscr{H}=\scrH_0+\scrH_z$).}

In the absence of SO coupling ($\phi=0$),
$\scrH_0$ has the exact SU(2) spin rotation symmetry
which is broken only ``weakly''~\cite{Oshikawa:2002aa}
by the Zeeman term $\scrH_Z$.
As a general principle of ESR, in this case, the ESR spectrum (if any)
remains paramagnetic, namely a single $\delta$-function at $\omega=H$.
As we will demonstrate later in Sec.~\ref{zeroorder},
in the low-temperature limit, this paramagnetic ESR can be
attributed to the edge states of the SSH model.
Once the SO coupling is introduced ($\phi \neq 0$),
the SU(2) symmetry is broken and we would expect 
a nontrivial ESR lineshape.
However, somewhat surprisingly, (as we will show later),
the ESR spectrum attributed to the edge states remains
a $\delta$-function at $\omega=H$ even when $\phi \neq 0$.

Thus, in order to investigate possible nontrivial effects of the SO
coupling on ESR, we further consider the NNN hoppings
\begin{equation}
\Delta\mathscr{H}=\sum_{j=1}\Delta
tc_{j+2}^\dagger\exp[i\gamma\vec{n}_\gamma\cdot\sigma/2]c_{j}+\text{h.c.} ,
\label{eq.NNN}
\end{equation}
where $\Delta t$ is the NNN electron hopping amplitude.
{The angle $\gamma$ and $\vec{n}_\gamma$ are the SO turn angle and 
the axis for the NNN hopping, respectively.}

The Hamiltonian of the system to be considered is then
\begin{eqnarray}
\label{separation}
\scrH_\text{ESR}&=&\scrH_0+ \scrH_Z + \Delta\scrH. 
\end{eqnarray}
Once we include the NNN hoppings,
the chiral symmetry is broken and the edge states are
not protected to be at zero energy.
Nevertheless, when the chiral symmetry breaking perturbations are weak,
we may still identify ``edge states'' localized near the ends
although they are no longer at exact zero energy even when
$H=0$.
Under a magnetic field $H$, contributions from these edge states dominate
the ESR in the low-energy limit.
Now, the ESR spectrum can be nontrivially modified by the
SO couplings $\phi$ and $\gamma$.
It is the main purpose of the present paper to elucidate this
effect.
In real materials, the NNN hoppings might be small but they are generally
non-vanishing.
Thus it is important to develop a theory of ESR in the presence
of the NNN hoppings, especially because we can detect
the NNN hoppings with ESR even when they are small.

We will treat the NNN hopping $\Delta \mathscr{H}$,
which would be smaller than the NN hopping $\mathscr{H}_0$
in many experimental realizations, as a perturbation.
This also turns out to be convenient for our theoretical analysis.
We also assume that {$|\phi|,|\gamma|\ll1$} since SO couplings are weak
in most of the realistic systems, and they will formulate a perturbation
expansion in $\Delta t$, $\phi$, and $\gamma$.

\section{Edge states of $\scrH_0$ and $U(1)_{\vec{S}\cdot\vec{n}_H}$ symmetry}
\label{zeroorder}
\label{edge}

As we discussed earlier, ESR would be an ideal probe
to detect the edge state of the 1D topological insulator
and various perturbations.
In this section, we discuss and explicitly solve the edge states
of the unperturbed Hamiltonian $\scrH_0$
to demonstrate the robustness of the edge states
against SO coupling.
As a consequence, in the model
$\mathscr{H}_0$ only with NN hoppings, there is no
nontrivial change in the edge ESR spectrum.


Since our Hamiltonian $\scrH_0$ is bilinear in fermion operators,
we can focus on single-electron states and represent them
with the ket notation.
For a half-infinite chain with sites $j=1,2,\ldots$, where $j=1$
corresponds to the end of the chain,
we find a single-electron eigenstate
$|\text{Edge},\sigma \rangle$
localized near the edge
in the ``topological insulator'' phase $\delta_0 >0$.
Here $\sigma=\pm 1$ represents the spin component in the
direction of the magnetic field.
Namely,
\begin{eqnarray}
\left( \scrH_0 + \scrH_Z \right) 
 |\text{Edge,}\sigma\rangle&=&E^{(0)}_\sigma|\text{Edge,}\sigma\rangle
\label{eq.energy_edge}
\\
 \vec{S}\cdot\vec{n}_H|\text{Edge,}\sigma\rangle&=&\frac{\sigma}{2}|\text{Edge,}\sigma\rangle,
\label{eq.spin_edge}
\end{eqnarray}
with the energy eigenvalues $E^{(0)}_\sigma=-\sigma H/2$ and $\vec{S}$ is the total spin of the system.
The wave function of the edge states is exactly calculated as
\begin{equation}
\label{zerothsolution}
\langle j, \sigma | \text{Edge,} \sigma' \rangle = \begin{cases}
\frac{\delta_{\sigma \sigma'}}{\sqrt{\mathscr{N}}} \left(-\frac{1-\delta_0}{1+\delta_0}\right)^{(j-1)/2}
& \mbox{ ($j \in 2 \mathbb{N}_0+1$ )}, \\
0 & (\mbox{otherwise}),
\end{cases}
\end{equation}
where $2 \mathbb{N}_0$ is the set of non-negative even integers, and
$\mathscr{N}$ is the normalization constant. 
The energy eigenvalue of the edge state for $\scrH_0$ is,
independently of the spin component, exactly zero,
reflecting its topological nature.

It is also remarkable that the edge state wavefunction is
independent of $\phi$ and $\vec{n}$.
This is a consequence of a canonical
``gauge transformation''~\cite{Starykh-SO-chains_PRB2008}
\begin{equation}
\begin{aligned}
\tilde{c}_{2k+1} &=c_{2k+1}, \\
\tilde{c}_{2k} &=\exp{\left(i\phi\vec{n}\cdot{\vec{\sigma}}/2\right)}c_{2k},
\end{aligned}
\label{Gtransformation}
\end{equation}
which eliminates the SO coupling from $\scrH_0$.
In this sense, $\scrH_0$ still has a hidden SU(2)
symmetry~\cite{Kaplan-SO_1983,SEA_PRL1992}
even though the SO coupling breaks the apparent spin SU(2) symmetry.
However, since the gauge transformation involves the
local rotation of spins,
the uniform magnetic field $\scrH_Z$ gives rise to
a staggered field after the gauge transformation.
This staggered field completely breaks the SU(2) symmetry.
This is similar to the situation in a spin chain with
a staggered Dzyaloshinskii-Moriya interaction~\cite{Oshikawa:1999aa}.
Thus, following the general principle of ESR, we would expect
a nontrivial ESR spectrum in the presence of the
staggered SO coupling as in $\scrH_0$.

Nevertheless, somewhat unexpectedly, we find that
the edge ESR spectrum for the model $\scrH_0$
remains the $\delta$-function $\delta(\omega-H)$,
as if there is no anisotropy at all.
This is due to the fact that the edge-state
wavefunction Eq.~\eqref{zerothsolution} is non-vanishing
only on the even sites.
Since the gauge transformation can be defined so that it
only acts on the even sites where the edge-state
wavefunction Eq.~\eqref{zerothsolution} vanishes,
the edge state is completely insensitive to the SO coupling.
{Therefore, the spectral shape of the edge ESR remains unchanged by the staggered SO coupling.} {In addition, since the edge wave functions are eigenstates of the total spin component along $\vec{n}_H$ according to Eq.~(\ref{eq.spin_edge}), the edge states have $U(1)_{\vec{S}\cdot\vec{n}_H}$ symmetry generated by $\vec{S}\cdot\vec{n}_H$. }

In fact, the robustness of the edge ESR spectrum is valid
for a wider class of models.
The edge ESR only probes a transition between two states
with opposite polarization of the spin, which form
a Kramers pair in the absence of the magnetic field.
Thus, at zero magnetic field, the time-reversal invariance
of the model requires these two states to be exactly degenerate. 
For a finite magnetic field, if the system still has $U(1)_{\vec{S}\cdot\vec{n}_H}$ symmetry of rotation
about the magnetic field axis, the two states can be
labelled by the eigenvalues of $S^z=\pm 1/2$, and
their energy splitting is exactly
\begin{eqnarray}
\omega_\text{ESR}=H.
\end{eqnarray}
Thus, as far as the edge ESR involving only the Kramers pair
is concerned, the $U(1)_{\vec{S}\cdot\vec{n}_H}$ symmetry is sufficient to protect
the $\delta$-function peak at $\omega =H$.
We note that, more generally, when more than two states contribute to ESR, these symmetries are not sufficient 
to protect the single-peak ESR spectrum, as there can be transitions between
states not related by time reversal.
In fact, this would be the case for the absorption
due to bulk excitations which we
do not discuss in this paper.

\section{Perturbation theory of the edge ESR frequency shift}
\label{sec:perturbation}

As we have shown in the previous Section, even in the presence
of SO coupling, there is no frequency shift for edge ESR
in the NN hopping model $\scrH_0$.
This is a consequence of the chiral symmetry, which stems from
the bipartite nature of the NN hopping model.

As we will discuss below, the introduction of
NNN hoppings breaks the chiral symmetry and generally
causes a nontrivial frequency shift of the edge ESR.
In this Section, we develop a perturbation theory of ESR for the edge states,
first by regarding $\scrH_0 + \scrH_Z$ as the unperturbed Hamiltonian and
$\Delta\mathscr{H}$ as a perturbation.

In the presence of the perturbation with SO couplings,
the eigenstate of the Hamiltonian is generally no longer
an eigenstate of the spin component $\vec{S}\cdot\vec{n}_H$
as in Eq.~\eqref{eq.spin_edge}.
Nevertheless, as long as the perturbation theory is valid,
{two edge states can still be identified by using spin $\sigma=+$ and $-$. 
Namely, for the full Hamiltonian~\eqref{separation}, 
we can define the edge state labeled by $\sigma=\pm$ as the state adiabatically 
connected to $|{\rm Edge},\pm\rangle$ as $\Delta\scrH \to 0$.}

{Here let us introduce new symbols $E_{+(-)}$ and $E_{+(-)}^{(n)}$ 
as the energy eigenvalue of the almost spin up (down) edge state 
in Eq.~\eqref{separation} and its $n$-th order correction in the perturbation theory, respectively.}
With these symbols, the ESR frequency is given by
\begin{equation}
 \omega_\text{ESR} = E_- - E_+ = H + \Delta \omega,
\end{equation}
where the ESR trivial peak position is given by $E^{(0)}_- - E^{(0)}_+=H$. 
The ESR frequency shift $\Delta \omega$, driven by the perturbation $\Delta\scrH$, 
is expanded in the perturbation theory as
\begin{equation}
 {\Delta \omega} = {\Delta \omega}^{(1)} + {\Delta \omega}^{(2)} + \ldots,
\end{equation}
where the $n$-th order term is
\begin{equation}
 {\Delta \omega}^{(n)} = E^{(n)}_- - E^{(n)}_+,
\label{eq.delta.omega_n}
\end{equation}
for $n \in \mathbb{N}$. In the following parts, we perturbatively solve the single-electron problem 
to compute the eigen-energy difference of two edge states $E_+$ and $E_-$.

\subsection{First order in $\Delta t$}
\label{firstorder}

The NNN hopping $\Delta \mathscr{H}$ breaks the chiral symmetry,
and thus it can change the edge ESR spectrum.
In fact, the energy of the edge states is already shifted
in the first order of $\Delta \mathscr{H}$.
However, the energy shift is the same for the two edge states
with different spin polarizations.
This is a consequence of the time-reversal (TR) symmetry of
the SO coupling, as demonstrated below:
\begin{eqnarray}
E_\sigma^{(1)}&=&\langle\text{Edge,}\sigma|\Delta\mathscr{H}|\text{Edge,}\sigma\rangle\nonumber\\
&=&\langle\Theta\left(\text{Edge,}-\sigma\right)|\Theta\Delta\mathscr{H}\Theta^{-1}|\Theta\left(\text{Edge,}-\sigma\right)\rangle\nonumber\\
&=&E_{-\sigma}^{(1)}\nonumber\\ &=&-2\Delta
t\frac{1-\delta_0}{1+\delta_0}\cos\left(\frac{\gamma}{2}\right)
\end{eqnarray}
where $\Theta$ is the TR operator and
$\Theta\Delta\mathscr{H}\Theta^{-1}=\Delta\mathscr{H}$ is
used.
Therefore, the edge ESR spectrum remains unchanged in the first order
of $\Delta t$ as the frequency shift vanishes in this order:
\begin{eqnarray}
\label{1st}
{\Delta \omega}^{(1)}=E^{(1)}_- -E^{(1)}_+ =0.
\end{eqnarray}

\subsection{Second order in $\Delta t$}
\label{sec:2nd}

We can formally write down the second-order perturbation correction
\begin{eqnarray}
E_\sigma^{(2)}=\langle\text{Edge,}\sigma|\Delta\mathscr{H}\!\!\cdot\!(E^{(0)}_{\sigma}-\mathscr{H}_0)^{-1}\!\!\cdot\!\!\mathscr{P}_\sigma\!\cdot\!\Delta\mathscr{H}|\text{Edge,}\sigma\rangle\nonumber
\end{eqnarray}
where the projection operator $\mathscr{P}_\sigma\equiv
1-|\text{Edge,}\sigma \rangle\langle\text{Edge,}\sigma|$. 
It is rather difficult to evaluate this formula directly,
since the intermediate states {in the perturbation term} 
include the bulk eigenstates of $\mathscr{H}_0$
in which the hidden symmetry is generally broken. 
Assuming that {$|\phi|,|\gamma| \ll 1$} 
(i.e., the SO coupling is sufficiently small),
we can develop a perturbative expansion in $\phi$ and
$\gamma$ in addition to $\Delta t$.
In this framework, we expand
$E_\sigma$ as a Taylor series of $\phi$, $\gamma$ and $\Delta t$.
The quantity of interest is the energy splitting
$E_- - E_+$, since it corresponds to the ESR frequency.

The edge state energy splitting in the second order in $\Delta t$,
and up to the second order in $\phi$, $\gamma$ 
is required to take the form
\begin{eqnarray}
 \label{2nd}
E^{(2)}_\sigma - E^{(2)}_{-\sigma} &\approx&
\frac{1}{2}\sigma H\Delta t^2\left\{a(\vec{n}_\gamma\times\vec{n}_H)^2\gamma^2+b(\vec{n}\times\vec{n}_H)^2\phi^2\right.\nonumber\\
&&\left.+c(\vec{n}\times\vec{n}_H)\cdot(\vec{n}_\gamma\times\vec{n}_H)\phi\gamma \right\},
\end{eqnarray}
based on the following symmetry considerations, {and $a$, $b$ and $c$ are constants to be determined}. 

First of all,
$(E_\sigma-E_{-\sigma})$ will not change under
\begin{eqnarray}
(H, \vec{n}_H,\sigma) \to (-H,-\vec{n}_H,-\sigma),
\end{eqnarray}
because this corresponds to a trivial redefinition of coordinate system. 
Therefore, we attach the factor $\sigma H$ in the r.h.s. of Eq.~(\ref{2nd}).

Next we notice that $\phi$ always appears with $\vec{n}$,
and $\gamma$ with $\vec{n}_\gamma$, which leads to further constraints as we will see below.
Let us consider the limiting case $\phi=0$ with nonzero $\gamma$.
The splitting can only depend on the relative angle between
$\vec{n}_\gamma$ and $\vec{n}_H$.
Furthermore, if $\vec{n}_\gamma \parallel \vec{n}_H$, the
hidden symmetry of the edge state implies that the energy
splitting is exactly given by the Zeeman energy
and there is no perturbative correction.
Thus, for $\phi=0$, the energy splitting can only
depend on $(\vec{n}_\gamma \times \vec{n}_H)^2 \gamma^2$
up to $O(\gamma^2)$ 
{since the energy split is a scalar and it must be written in terms
of inner and vector products of $\vec{n}_H$, $\vec{n}$ and $\vec{n}_\gamma$.} 
Then, similarly, if $\gamma=0$ with nonzero $\phi$, the energy splitting can only
depend on $(\vec{n} \times \vec{n}_H)^2 \phi^2$.
Finally, the $O(\phi \gamma)$ term should be linear in
$\vec{n}$ and $\vec{n}_\gamma$, and it vanishes when
$\vec{n} \parallel \vec{n}_\gamma \parallel \vec{n}_H$
because of the $U(1)_{\vec{S}\cdot\vec{n}}$ symmetry.
These requirements uniquely {determine} the form of $(\vec{n}\times\vec{n}_H)\cdot(\vec{n}_\gamma\times\vec{n}_H)$.
Thus, the symmetries reduce the possible forms
of the second-order corrections to Eq.~\eqref{2nd} with
only the three parameters $a$, $b$, and $c$.

{To obtain these parameters,} we note that the expansion in $\phi$ and $\gamma$
introduced above can be naturally done by
regarding the SSH model without SO coupling 
\begin{eqnarray}
\label{unperturbh}
\tilde{\mathscr{H}}_0&=&-\sum_{j=1}^{+\infty}\left\{t\left[1+(-1)^j\delta_0\right]c_{j+1}^\dagger c_j+\text{h.c.}\right\}\nonumber\\
&&-H\sum_{j=1}c^\dagger_j\left(\vec{\sigma}\cdot\vec{n}_H\right)c_j/2
\label{pertHamiltonian}
\end{eqnarray}
as the unperturbed Hamiltonian, and
\begin{equation}
\mathscr{H}_\text{pert} = \Delta\mathscr{H}_0+\Delta\mathscr{H}, 
\label{eq.perturbation}
\end{equation}
as the perturbation, where
\begin{eqnarray}
\Delta\mathscr{H}_0&\equiv&-\sum_{j=1}^{+\infty}\left\{t\left[1+(-1)^j\delta_0\right]c_{j+1}^\dagger\right.\nonumber\\
&&\left.\left\{\exp\left[(-1)^ji\phi\vec{n}\cdot\vec{\sigma}/2\right]-1\right\}c_j+\text{h.c.}\right\}
\end{eqnarray}
and $\Delta\mathscr{H}$ as defined in Eq.~\eqref{eq.NNN}.

{As the result of the perturbation calculation given in the Appendix,} we find that the second-order term of frequency shift is non-positive given in the form
\begin{eqnarray}
\label{energyshift}
\Delta \omega^{{(2)}} &=&
 -\frac{H}{2}\sum_{m_3,j_3}(\vec{M}+\vec{N})^\dagger\cdot(\vec{M}+\vec{N})
\leq 0,
\end{eqnarray}
where
\begin{eqnarray}
\vec{M}&\equiv&\sum_{m_2,j_2}\!\!\frac{\langle m_3,j_3|\Delta\mathscr{H}'_0|m_2,j_2\rangle\langle m_2,j_2|\Delta\mathscr{H}''|\text{Edge}\rangle}{m_2E_{j_2}E_{j_3}}\nonumber\\
&&(\vec{n}\times\vec{n}_H)\phi\nonumber\\
\vec{N}&\equiv&\frac{\langle m_3,j_3|\Delta\mathscr{H}'|\text{Edge}\rangle}{E_{j_3}}(\vec{n}_\gamma\times\vec{n}_H)\gamma.
\end{eqnarray}
Here $|m,j\rangle$'s are single-particle (bulk) energy eigenstates of $\tilde{\mathscr{H}}_0$ in Eq.~(\ref{unperturbh}) where $m=\pm$ labels the positive or negative energy sector (i.e., band indices) and $j$ labels other possible quantum numbers, which is not the wave vector since we have the open-ended boundary condition. 
The energy $\pm E_j$ stands for the energy eigenvalue of $|\pm,j\rangle$. 
{The perturbation terms 
$\Delta\mathscr{H}'$, $\Delta\mathscr{H}''$, and $\Delta\mathscr{H}'_0$ are defined as}
\begin{eqnarray}
\label{deltah'}
&&\Delta\mathscr{H}'\equiv\sum_{j=1}\Delta tc^\dagger_{j+2}c_j-\text{h.c.}, \\
\label{deltah''}
&&\Delta\mathscr{H}''\equiv\sum_{j=1}\Delta tc^\dagger_{j+2}c_j+\text{h.c.}, \\
&&\Delta\mathscr{H}'_0\equiv-\sum_{j=1}^{+\infty}\left\{t\left[1+(-1)^j\delta_0\right]c_{j+1}^\dagger c_j-\text{h.c.}\right\} .
\label{deltah0'}
\end{eqnarray}
We note that Eqs.~\eqref{deltah'} and~\eqref{deltah0'} are anti-Hermitian.

After putting the definitions of $\vec{M}$ and $\vec{N}$ into
Eq.~(\ref{energyshift}), we find the result consistent with
the general form Eq.~(\ref{2nd}) required by symmetries.
The parameters are then identified as
\begin{eqnarray}
a&=&b=\left[\frac{\Delta t}{t(1+\delta_0)}\right]^2,\quad
c=-2\left[\frac{\Delta t}{t(1+\delta_0)}\right]^2. \nonumber
\end{eqnarray}
The detailed derivation of $a$, $b$ and $c$ is given in Appendix.
The final result can be given in a compact form as
\begin{equation}
{\Delta \omega}^{(2)} =
-\frac{H}{2}\left|
\frac{\Delta t}{t(1+\delta_0)}(\phi\vec{n}\times\vec{n}_H-\gamma\vec{n}_\gamma\times\vec{n}_H)\right|^2,
\label{peakshift}
\end{equation}
which is non-positive.
Since the first-order correction vanishes as we have already seen,
the second-order term Eq.~\eqref{peakshift} gives the leading term for the
frequency shift of the edge ESR.
It also implies that, although $\gamma$ is
the SO-coupling turn angle for the NNN hopping terms, it is equally important in
$\Delta \omega$ as the NN SO coupling turn angle $\phi$
even if the NNN hopping itself is small ($\Delta t \ll t$).

One of the most remarkable features of our result is that,
the shift (up to the second order in the perturbation) vanishes
when $\phi\vec{n}\times\vec{n}_H=\gamma\vec{n}_\gamma\times\vec{n}_H$.
This corresponds to the zeros of curves
in Fig.~\ref{figpeakshift} where the direction of the magnetic field is in 
the plane spanned by $\vec{n}=\hat{y}$ and $\vec{n}_\gamma=-\hat{z}$, and $\theta$ is the angle between $\vec{n}_H$ and $\hat{y}$ in the $\hat{y}$-$\hat{z}$ plane, as shown in Fig.~\ref{coordinate}. Therefore, we predict two directions of the magnetic field
for which the ESR shift vanishes,
when the magnetic field direction $\vec{n}_H$ sweeps the plane spanned
by $\vec{n}$ and $\vec{n}_\gamma$.

\section{Non-perturbative calculation of the edge ESR spectrum}
\label{sec:numerical}

Here, in order to see the validity of our perturbation theory in the preceding section, 
let us re-compute the edge ESR frequency with a more direct numerical method.
As already mentioned, when the magnetic field $H$ (and thus the ESR frequency $\omega$)
is much smaller compared to the bulk excitation gap $4t|\delta_0|$, we may ignore the
effects of the bulk excitations in the ESR spectrum.
Then the edge ESR spectrum is of the $\delta$-function form
\begin{eqnarray}
\chi''_{+-\text{edge}}(q=0,\omega)&\propto&\delta\left[\omega-(E_--E_+)\right],
\end{eqnarray}
where $E_{+(-)}$ is the energy eigenvalue of the ``almost'' spin-up (spin-down) edge state. 
These energies can be accurately computed by numerical diagonalization of a
finite (but long) size full Hamiltonian with an open boundary condition. 
Then we obtain the spectrum peak shift ${\Delta \omega} \equiv E_- -E_+ -H$ 
from the numerical results of $E_\pm$. 
In the present work, we calculated $E_\pm$ using a finite open chain of 100 sites.

\begin{figure}
  \centering
\includegraphics[width=3.3in]{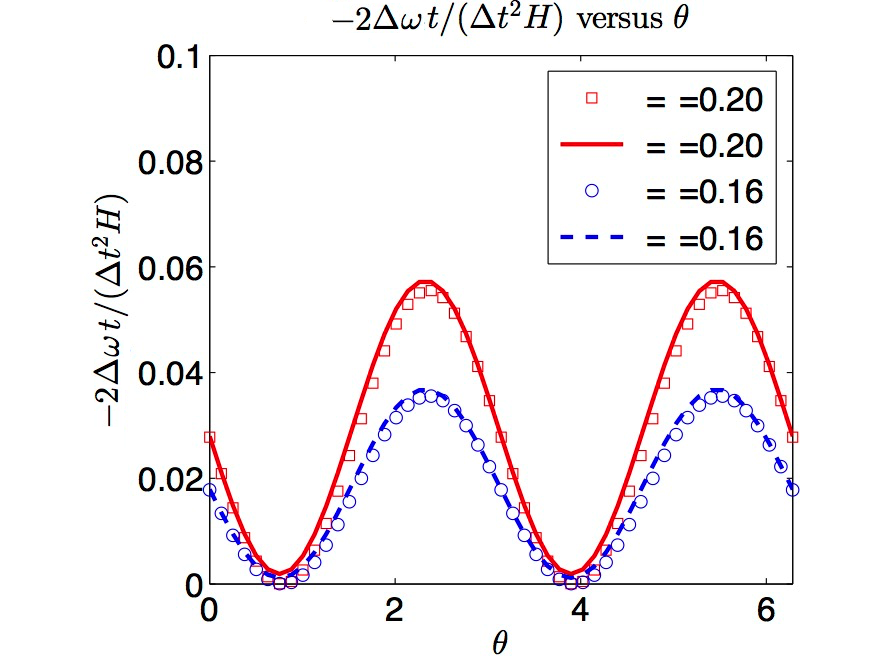}
  \caption{Angle $\theta$ dependence of the ESR frequency shift 
$\Delta \omega$. We set the parameters $t=1.0$, $\Delta t=\delta_0=0.2$, 
and $H=0.05$. Red and blue curves are obtained by the second-order 
perturbation calculation in Sec.~\ref{sec:2nd}. Square and circle points are 
the results of direct numerical diagonalization for a system (of 100 sites)
with an open boundary condition. The definition of angle $\theta$ is
indicated in Fig.~\ref{coordinate}. The zeros of $\Delta \omega$ occur at 
$\phi\vec{n}\times\vec{n}_H=\gamma\vec{n}_\gamma\times\vec{n}_H$.
}
\label{figpeakshift}
 \end{figure}

\begin{figure}
  \centering
 \includegraphics[width=3.3in]{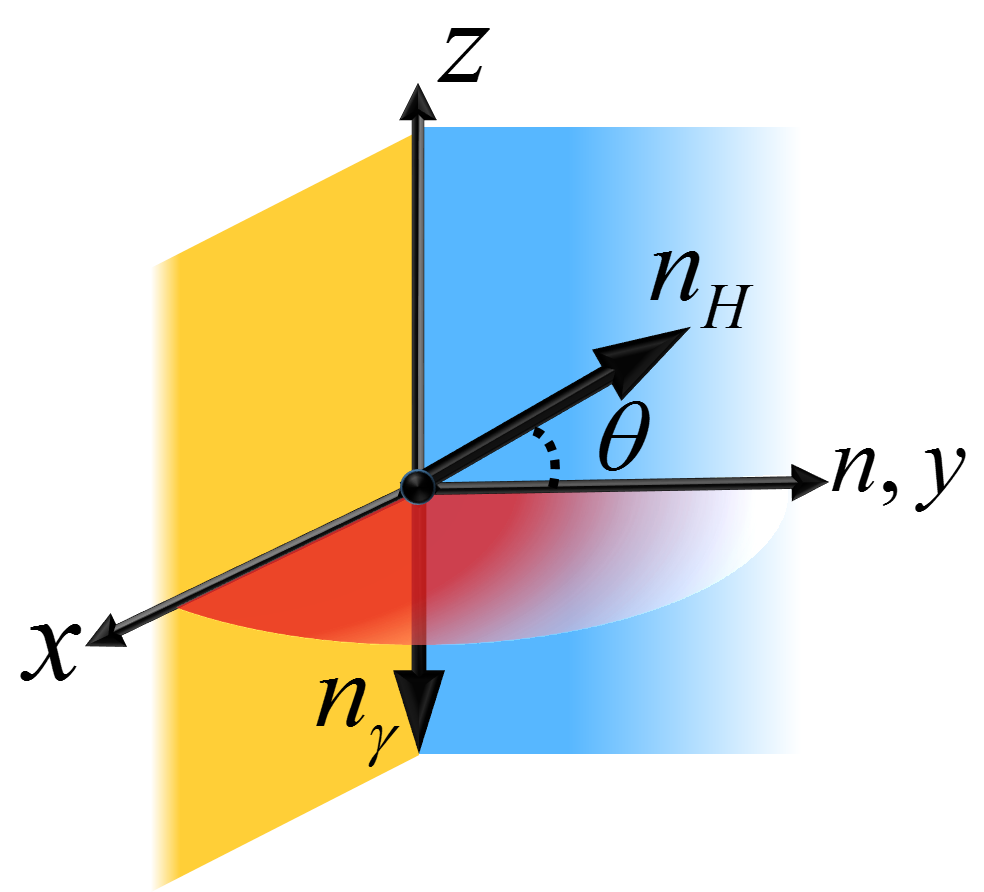}
  \caption{{Geometric relation among some vectors in Fig.~\ref{figpeakshift}, 
where the direction of magnetic field is in the plane spanned by $\vec{n}=\hat{y}$ and $\vec{n}_\gamma=-\hat{z}$. 
The parameter $\theta$ is defined as 
the angle between $\vec{n}_H$ and $\hat{y}$ in the $\hat{y}$-$\hat{z}$ plane.}
}
\label{coordinate}
 \end{figure}

In Fig.~\ref{figpeakshift}, we compare the numerical results of the
peak shift with the analytical perturbation theory of
$\Delta\omega_\text{ESR}^{(2)}$ in Eq.~(\ref{peakshift}) when the magnetic field
is in the plane spanned by $\vec{n}$ and $\vec{n}_\gamma$. 
The figure clearly shows that our perturbation theory agrees with the numerical results quite well.

\section{Conclusions and Discussion}
\label{conclusion}

We have analyzed ESR of edge states in a generalized
SSH model with staggered SO couplings and with an open end.
In this paper, we assume that
the energy scales of the magnetic field, the frequency, and the temperature
are sufficiently small compared to the bulk gap.
Then the ESR spectrum only consists of a single $\delta$-function
spectrum corresponding to the transition between two spin states
at the edge, but is expected to show a nontrivial frequency shift in general
as the SO coupling breaks the SU(2) symmetry strongly under the
applied magnetic field.
Nevertheless, there is no ESR frequency shift in the model
with only NN hoppings, thanks to its chiral symmetry.

The chiral symmetry is broken by NNN hoppings, which should be
generally present in any realistic materials even if they are small.
This NNN hoppings, together with the SO coupling, can induce
a nontrivial frequency shift on the edge ESR.
Thus we have developed a perturbation theory of the frequency
shift, regarding the NNN hoppings and the SO couplings as perturbations.
Our main result, the ESR frequency shift up to second order in the
perturbation theory, is found in Eq.~\eqref{2nd}.
It is non-positive in this order.
(The resonance field shift for a fixed frequency, which is usually
measured in experiments, is always positive.)
In the presence of the NNN hoppings, the SO couplings in the
NN hoppings, which did not cause a frequency shift by themselves,
also contribute to the frequency shift.
We find an interesting dependence of the ESR
frequency shift on the direction of the static magnetic field,
relative to the SO couplings on the NN and the NNN hoppings.
In particular, the ESR frequency shift is predicted to vanish when
the static magnetic field points to a certain direction
on the plane spanned by the two SO coupling axes (see Fig.~\ref{figpeakshift}).
Furthermore, we performed a direct estimate of the ESR frequency
shift by a numerical calculation of the edge state spectrum,
without relying on the perturbation theory.
The result agrees very well with the perturbation theory,
establishing its validity.

Our results indicate that, the chiral symmetry breaking by the NNN
hoppings in the ``SSH''-type topological insulators in one dimension
may be detected by ESR, in the presence of the SO couplings.
If the NNN hoppings are small, which would be the case in many
realistic materials, it might be difficult to detect their effects
with other experimental techniques.
ESR has been successful in detecting even very small magnetic
anisotropies, thanks to its high sensitivity and accuracy.
We hope that the present work will pave the way for a new application of ESR
in detecting (small) chiral symmetry breaking.

While we do not discuss any particular material in this paper,
let us discuss here the prospect of experimentally observing
the effects we predict.
The maximal frequency shift is given by the order of
\begin{equation}
 H \left(\frac{\Delta t}{t} \max{ \{\phi, \gamma\} } \right)^2 .
\end{equation}
The ratio of NNN to NN hoppings, $\Delta t/t$, of course
strongly depends on each material.
It is even possible that $|\Delta t|/ t \gg 1$, in which case
the system may be regarded as two chains coupled weakly
by zigzag hopping.
It should be however noted that our theory is valid only when $|\Delta t|/ t$
is sufficiently small.
We still expect that our theory works reasonably well for
$(\Delta t /t)^2 \sim 0.1$.
In carbon-based systems, such as polyacetylene, the
SO interaction is known to be weak.
For example, even with the enhanced SO interaction
due to a curvature~\cite{Huertas-Hernando_SO-curved_PRB2006},
$\phi, \gamma$ is of order of $10^{-5}$.
This would give an ESR shift that is too small to be observed in experiments.
However, the SO interaction is stronger in heavier atoms.
In fact, even in carbon-based systems,
the SO interaction can be significantly enhanced
by heavy adatoms.
For example, placing Pb as adatoms can enhance $\phi, \gamma$
up to $0.1$ or more in graphene~\cite{Brey_graphene-SO-adatoms_PRB2015}.
This would give the edge ESR shift corresponding to the
$g$-shift up to the order of $10^{-3}$ (1,000 ppm), which
should be observable.
In particular, even if the absolute value of the shift is difficult
to be determined, the angular dependence of the ESR shift
would be more evident in experiments.
{Furthermore, it is known that an SO interaction generally becomes larger 
when the electron system we consider is located in the vicinity of an interface between two bulk systems 
or is under a strong, static electric field \cite{Nitta:1997aa,Caviglia:2010aa,Soumyanarayanan:2016aa}. Therefore, if we set up an SSH chain system under such an environment, 
it would become easier to detect an ESR frequency shift due to a strong SO coupling.}

Throughout this paper, we have taken the NN SO coupling in the model
Eq.~(\ref{H0}) to be staggered.  However, there are other possibilities.
In particular, in a translationally symmetric system, the NN SO coupling
is uniform. Although the only difference is the signs in the
Hamiltonian, ESR spectra should be significantly different between
these two cases.
This is clear if we consider the limit of the zero NNN hopping.
In the staggered SO coupling
case, there is no ESR frequency shift. 
This is because the NN SO coupling can be gauged out
by the canonical transformation Eq.~(\ref{Gtransformation}),
without changing the odd site amplitude and
thus leaving the SSH edge state wavefunction Eq.~(\ref{zerothsolution}) unchanged.
On the other hand, in the uniform SO coupling case,
gauging out the NN SO coupling {affects}
any wavefunction including the SSH edge state wavefunction,
resulting in the change of the ESR spectrum.
A similar difference has been recognized between the ESR spectum
in the presence of a staggered DM interaction~\cite{Oshikawa:1999aa}
and that with a uniform DM interaction along
the chain~\cite{Starykh-SO-chains_PRB2008}.
The analysis of the edge ESR spectrum in the presence
of a uniform SO coupling is left for future studies.

\begin{acknowledgments}
M. S. was supported by
Grant-in-Aid for Scientific Research on Innovative Area,
``Nano Spin Conversion Science'' (Grant No.17H05174),
and JSPS KAKENHI Grants (No. 17K05513 and No. 15H02117),
and M. O. by KAKENHI Grant No. 15H02113.
This work was also supported in part by
U.S. National Science Foundation under Grant No. NSF PHY-1125915
through Kavli Institute for Theoretical Physics, UC Santa Barbara
where a part of this work was performed by Y. Y. and M. O.

\end{acknowledgments}

\appendix*
\section{Perturbation theory for the frequency shift $\Delta\omega$}
\label{appendix}

Here we explain how to determine the parameters 
$a$, $b$ and $c$ in the general form of Eq.~(\ref{2nd}), based on a perturbation theory. 
As we have discussed earlier, in order to expand the edge-state energy eigenstates 
with respect to the SO coupling parameters $\phi$ and $\gamma$,
which are assumed to be small, we regard $\tilde{\scrH}_0$ of Eq.~\eqref{pertHamiltonian} 
as the unperturbed part, and $\tilde{\scrH}_\text{pert}$ of Eq.~\eqref{eq.perturbation} as the perturbation.
Moreover, we have already shown in Sections~\ref{zeroorder}
and~\ref{firstorder} that the ESR shift vanishes exactly up to $O(\Delta t)$. 
Therefore, the leading terms with coefficients $a$, $b$ and $c$ stem from the second-order perturbation 
proportional to $\Delta t^2$. Below, we will determine three parameters $a$, $b$ and $c$ 
from the perturbative expansion of $\tilde{\scrH}_\text{pert}$.

\subsection{Second order in $\scrH_\text{pert}$}

The shift of the edge energy eigenvalue in the second order of $\scrH_\text{pert}$ is given by 
\begin{eqnarray}
\label{eq:E2}
E^{[2]}_\sigma\!\!&=&\!\!\!\!\!\sum_{m=\pm,j,\tilde{\sigma}=\pm}\!\!\!\!\!\!\!\!\frac{\langle\text{Edge,}\sigma|\mathscr{H}_\text{pert}|m,j,\tilde{\sigma}\rangle\!\langle m,j,\tilde{\sigma}|\mathscr{H}_\text{pert}|\text{Edge,}\sigma\rangle}{-\sigma\frac{H}{2}-E_{m,j,\tilde{\sigma}}}\nonumber\\
&\approx&\sum_{m=\pm,j}\frac{\langle\text{Edge}|(\Delta\mathscr{H}')^\dagger|m,j\rangle\langle m,j|\Delta\mathscr{H}'|\text{Edge}\rangle}{E^2_{j}}\sigma H\nonumber\\
&&\cdot|n_\gamma\times n_H|^2{\gamma^2}/{4},
\end{eqnarray}
where the bulk single-particle eigen-energy $E_{m,j,\tilde{\sigma}}$ is given by 
\begin{eqnarray}
E_{m,j,\tilde{\sigma}}&\equiv&mE_{j}-\tilde{\sigma}{H}/{2}
\end{eqnarray}
with $mE_j$ the energy eigenvalue of spinless single-particle eigenstates $|m,j\rangle$, and we have used the facts that in the unperturbed sector, the orbital and spin parts of single-particle eigenstates 
can be decomposed, e.g. 
$|m,j,\sigma\rangle=|m,j\rangle|\sigma\rangle$ since $\tilde{\mathscr{H}}_0$ commutes 
with the spin operator of each site. 
Here we define $E^{[n]}_\sigma$ as the energy eigenvalue of the edge state
with spin $\sigma$ in the $n$-th order of perturbation in $\scrH_\text{pert}$.
This is to be distinguished from $E^{(n)}_\sigma$ introduced
in Eq.~\eqref{eq.delta.omega_n}, where $n$ refers to the order in $\Delta t$.
In the second line of Eq.~(\ref{eq:E2}), we assume $|E_{m,j,\tilde{\sigma}}|\gg H$ and 
perform the Taylor expansion of $E^{[2]}_\sigma$ with respect to $H$. 
The matrix $\langle m,j|\Delta\mathscr{H}'|\text{Edge}\rangle$ can be calculated by using 
the anti-Hermitian operator $\Delta\mathscr{H}'$ of Eq.~(\ref{deltah'}).  
From Eq.~(\ref{eq:E2}), the ESR frequency shift driven by $E^{(2)}_\sigma$ is expressed as  
\begin{eqnarray}
\label{Appendix-2nd}
&&E_-^{[2]}-E_+^{[2]}=-|n_\gamma\times n_H|^2{\gamma^2}/{2}\nonumber\\
&&\!\!\cdot\sum_{m=\pm,j}\!\!\!\frac{\langle\text{Edge}|(\Delta\mathscr{H}')^\dagger|m,j\rangle\langle m,j|\Delta\mathscr{H}'|\text{Edge}\rangle}{E^2_{j}}H.
\end{eqnarray}
This indeed corresponds to the parameter $a$. 
To obtain $b$ and $c$, we should proceed to higher orders of $\scrH_\text{pert}$.

\subsection{Third order in $\scrH_\text{pert}$}
In the third order perturbation theory within $\Delta t^2$, the energy correction takes the form as
\begin{widetext}
\begin{eqnarray}
&&E_\sigma^{[3]}=\sum_{m,j,\tilde{\sigma}}\frac{\langle\text{Edge,}\sigma|\mathscr{H}_\text{pert}|m_3,j_3,\tilde{\sigma}_3\rangle\langle m_3,j_3,\tilde{\sigma}_3|\mathscr{H}_\text{pert}|m_2,j_2,\tilde{\sigma}_2\rangle\langle m_2,j_2,\tilde{\sigma}_2|\mathscr{H}_\text{pert}|\text{Edge,}\sigma\rangle}{\left[-(\sigma-\tilde{\sigma}_2)\frac{H}{2}-m_2E_{j_2}\right]\left[-(\sigma-\tilde{\sigma}_3)\frac{H}{2}-m_3E_{j_3}\right]}\nonumber\\
&&+\langle\text{Edge,}\sigma|\mathscr{H}_\text{pert}|\text{Edge,}\sigma\rangle\frac{|\langle\text{Edge,}\sigma|\mathscr{H}_\text{pert}|m_3,j_3,\tilde{\sigma}_3\rangle|^2}{\left[-(\sigma-\tilde{\sigma}_3)\frac{H}{2}-m_3E_{j_3}\right]^2}\nonumber\\
&\approx&\sum_{m,j}\text{Re}\left[\frac{\langle\text{Edge}|\Delta\mathscr{H}''|m_3,j_3\rangle\langle m_3,j_3|\Delta\mathscr{H}'_0|m_2,j_2\rangle\langle m_2,j_2|\Delta\mathscr{H}'|\text{Edge}\rangle}{m_3E^2_{j_2}E_{j_3}}\sigma H\right]\cdot(\vec{n}\times\vec{n}_H)\cdot(\vec{n}_\gamma\times\vec{n}_H)\frac{\phi\gamma}{2},
\end{eqnarray}
\end{widetext}
where $\Delta\mathscr{H}''$ is defined by Eq.~(\ref{deltah''}). 

Therefore,
\begin{eqnarray}
\label{3rd}
& &E_-^{[3]}-E_+^{[3]}\nonumber\\
&=&\sum_{m,j}\!-\!\text{Re}\!\!\left[\frac{\langle\text{Edge}|\Delta\mathscr{H}''|m_3,j_3\rangle\langle m_3,j_3|(\Delta\mathscr{H}'_0)^\dagger|m_2,j_2\rangle}{m_3E^2_{j_2}E_{j_3}}\right.\nonumber\\
&&\left.\cdot\langle m_2,j_2|\Delta\mathscr{H}'|\text{Edge}\rangle H\right](\vec{n}\times\vec{n}_H)\cdot(\vec{n}_\gamma\times\vec{n}_H)\phi\gamma. \nonumber\\
\end{eqnarray}
We see that this correction term corresponds to the parameter $c$ in Eq.~(\ref{2nd}).

\subsection{Fourth order in $\scrH_\text{pert}$}

In order to derive the leading term of the parameter $b$, we have to calculate the fourth-order term. 
For convenience of the fourth-order calculation, we introduce the abbreviated notation of the matrix elements:
\begin{eqnarray}
A^{qr}&\equiv&\langle m_r,j_r,\tilde{\sigma}_r|A|m_q,j_q,\tilde{\sigma}_q\rangle, \\
E_{\text{E}q}&\equiv&E^{[0]}_\sigma-E_{m_q,j_q,\tilde{\sigma}_q}
\end{eqnarray}
for any operator $A$, and denote the zeroth order edge state by ``E''. 
In this notation, the fourth order edge-energy correction up till $(\Delta t\phi)^2$-order is  given by 
\begin{widetext}
\begin{eqnarray}
\label{fourthorder}
&&E_\sigma^{[4]}=\sum_{m,j,\tilde{\sigma}}\frac{\Delta\mathscr{H}^{\text{E}4}\mathscr{H}_\text{pert}^{4,3}\mathscr{H}_\text{pert}^{3,2}\Delta\mathscr{H}^{2\text{E}}}{E_{\text{E}2}E_{\text{E}3}E_{\text{E}4}}-E_\sigma^{[2]}\frac{\left(\Delta\mathscr{H}^{\text{E}4}\right)^2}{(E_{\text{E}4})^2}-2\Delta\mathscr{H}^\text{EE}\frac{\Delta\mathscr{H}^{\text{E}4}\mathscr{H}_\text{pert}^{43}\Delta\mathscr{H}^{3\text{E}}}{E_{\text{E}3}^2E_{\text{E}4}}+\left(\Delta\mathscr{H}^\text{EE}\right)^2\frac{\left(\Delta\mathscr{H}^{\text{E}4}\right)^2}{(E_{\text{E}3})^3}\nonumber\\
&\approx&\sum_{m,j,}\frac{\langle\text{Edge}|\Delta\mathscr{H}''|m_2,j_2\rangle\langle m_2,j_2|(\Delta\mathscr{H}'_0)^\dagger|m_3,j_3\rangle\langle m_3,j_3|\Delta\mathscr{H}'_0|m_4,j_4\rangle\langle m_4,j_4|\Delta\mathscr{H}''|\text{Edge}\rangle}{m_2m_4E_{j_2}E^2_{j_3}E_{j_4}}{\sigma H}|\vec{n}\times\vec{n}_H|^2\frac{\phi^2}{4}.
\end{eqnarray}
Then we can arrive at 
\begin{eqnarray}
\label{4th}
&&E_-^{[4]}-E_+^{[4]}\nonumber\\
&\approx&\!-\!\!\sum_{m,j,}\!\!\frac{\langle\text{Edge}|\Delta\mathscr{H}''|m_2,j_2\rangle\langle m_2,j_2|(\Delta\mathscr{H}'_0)^\dagger|m_3,j_3\rangle\langle m_3,j_3|\Delta\mathscr{H}'_0|m_4,j_4\rangle\langle m_4,j_4|\Delta\mathscr{H}''|\text{Edge}\rangle}{m_2m_4E_{j_2}E^2_{j_3}E_{j_4}}{H}|\vec{n}\times\vec{n}_H|^2\frac{\phi^2}{2}.
\end{eqnarray}
\end{widetext}
This correspond to the term coefficiented by the parameter $b$ in  Eq.~(\ref{2nd}).

\subsection{Summation over the perturbation orders}
Through some algebra, we find that terms of the fifth and higher orders in $\scrH_\text{pert}$ do not contribute 
to the order of $\mathscr{O}\left(\Delta t^2\phi^2\right)$. 
Summing up Eqs.~(\ref{1st}), (\ref{Appendix-2nd}), (\ref{3rd}), and (\ref{4th}), 
we arrive at the energy shift of the energy difference between down- and up-spin edge states 
in an elegant form as in Eq.~\eqref{energyshift}.

%

\end{document}